\documentclass[letterpaper, 10 pt, conference]{ieeeconf}

\IEEEoverridecommandlockouts                              

\usepackage{graphicx}      
\usepackage{amssymb,amsmath}
\usepackage{algorithm}
\usepackage{algpseudocode}
\usepackage{subfigure}
\usepackage{epstopdf}   
\usepackage{tikz}
\usetikzlibrary{arrows,matrix,positioning,fit,calc}
\usepackage{nicematrix}

\usepackage{fancyhdr}

\makeatletter 
\let\myorg@bibitem\bibitem
\def\bibitem#1#2\par{%
  \@ifundefined{bibitem@#1}{%
    \myorg@bibitem{#1}#2\par
  }{%
    \begingroup
      \color{\csname bibitem@#1\endcsname}%
      \myorg@bibitem{#1}#2\par
    \endgroup
  }%
}
\newcommand*{\bibitem@aboudonia}{black}    
\newcommand*{\bibitem@madistributed}{black}    
\newcommand*{\bibitem@riverso}{black}    
\newcommand*{\bibitem@tucci}{black}    
\newcommand*{\bibitem@shalmani}{black}    

\makeatother 

\newcommand{\tikzmark}[1]{\tikz[overlay,remember picture] \node (#1) {};}
\newcommand{\DrawBox}[4][]{%
    \tikz[overlay,remember picture]{%
        \coordinate (TopLeft)     at ($(#2)+(-0.3,0.95em)$);
        \coordinate (BottomRight) at ($(#3)+(0.6 em,-0.1em)$);
        \coordinate (LabelPoint) at ($(TopLeft)!0.5!(BottomRight)$);
        \draw [black,#1] (TopLeft) rectangle (BottomRight);
        \node [ #1, fill=none, fill opacity=1] at (LabelPoint) {#4};
    }
}
\newcommand{\DrawBoxTwo}[4][]{%
    \tikz[overlay,remember picture]{%
        \coordinate (TopLeft)     at ($(#2)+( -0,0.9em)$);
        \coordinate (BottomRight) at ($(#3)+(0.1 em, -0.3em)$);
        \coordinate (LabelPoint) at ($(TopLeft)!0.5!(BottomRight)$);
        \draw [black,#1] (TopLeft) rectangle (BottomRight);
        \node [ #1, fill=none, fill opacity=1] at (LabelPoint) {\small#4};
    }
}
\newcommand{\DrawBoxThree}[4][]{%
    \tikz[overlay,remember picture]{%
        \coordinate (TopLeft)     at ($(#2)+( -0.33,1.2em)$);
        \coordinate (BottomRight) at ($(#3)+(1.1 em, -0.1em)$);
        \coordinate (LabelPoint) at ($(TopLeft)!0.5!(BottomRight)$);
        \draw [black,#1] (TopLeft) rectangle (BottomRight);
        \node [ #1, fill=none, fill opacity=1] at (LabelPoint) {#4};
    }
}
\newcommand{\DrawBoxFive}[4][]{%
    \tikz[overlay,remember picture]{%
        \coordinate (TopLeft)     at ($(#2)+( -0,1.0em)$);
        \coordinate (BottomRight) at ($(#3)+(0.1 em, -0.3em)$);
        \coordinate (LabelPoint) at ($(TopLeft)!0.5!(BottomRight)$);
        \draw [black,#1] (TopLeft) rectangle (BottomRight);
        \node [ #1, fill=none, fill opacity=1] at (LabelPoint) {#4};
    }
}
\newcommand{\DrawBoxFour}[4][]{%
    \tikz[overlay,remember picture]{%
        \coordinate (TopLeft)     at ($(#2)+( -0,0.9em)$);
        \coordinate (BottomRight) at ($(#3)+(0.1 em, -0.5em)$);
        \coordinate (LabelPoint) at ($(TopLeft)!0.5!(BottomRight)$);
        \draw [black,#1] (TopLeft) rectangle (BottomRight);
        \node [ #1, fill=none, fill opacity=1] at (LabelPoint) {\small#4};
    }
}

\newcommand{\DrawBoxSix}[4][]{%
    \tikz[overlay,remember picture]{%
        \coordinate (TopLeft)     at ($(#2)+(-0.4,0.95em)$);
        \coordinate (BottomRight) at ($(#3)+(1.0 em,-0.1em)$);
        \coordinate (LabelPoint) at ($(TopLeft)!0.5!(BottomRight)$);
        \draw [black,#1] (TopLeft) rectangle (BottomRight);
        \node [ #1, fill=none, fill opacity=1] at (LabelPoint) {#4};
    }
}

\newtheorem{pro}{Proposition}

\usepackage{float}
\usepackage{lmodern}
\usepackage{subfig}

\usepackage{url}

\usepackage{boldline} 

\usepackage{xcolor} %

\usepackage{calrsfs}
\allowdisplaybreaks

\DeclareMathAlphabet{\pazocal}{OMS}{zplm}{m}{n}

\DeclareMathOperator*{\argminA}{arg\,min}
\newcommand{\norm}[1]{\left\lVert#1\right\rVert}

\usepackage{color,soul}

\newcommand{\resp}[1]{{\color{black}  #1}}

\title{\LARGE \bf \resp{A distributed framework for linear adaptive MPC}}
\author{Anilkumar Parsi$^*$, Ahmed Aboudonia$^*$, Andrea Iannelli, John Lygeros, and Roy S. Smith
\thanks{
\noindent This work has been partially supported by the Swiss National Science Foundation under grant no. $200021\_178890$.
\newline
The authors are with the Department of Electrical Engineering, Automatic Control Lab, ETH, Z\"{u}rich 8092, Switzerland
{\tt\footnotesize aparsi/ahmedab/iannelli/lygeros/rsmith@control.ee.ethz.ch}.
\newline
$^*$Both the authors contributed equally to this work.\vspace*{6pt}
\newline
\copyright\ 2021 IEEE.  Personal use of this material is permitted.  Permission from IEEE must be obtained for all other uses, in any current or future media, including reprinting/republishing this material for advertising or promotional purposes, creating new collective works, for resale or redistribution to servers or lists, or reuse of any copyrighted component of this work in other works.}}

\begin{document}

\maketitle


\thispagestyle{fancy}
\renewcommand{\headrulewidth}{0pt}
\setlength{\headheight}{24pt}
\chead{\color{red} \framebox{\begin{minipage}{\textwidth}Published: \emph{Proc.\ 60$^{\text{th}}$ IEEE
Conf.\ Decision \& Control,} pp.~460--465, 2004. DOI: 10.1109/CDC45484.2021.9683290
\end{minipage} }}
\cfoot{}
\pagestyle{empty}

\begin{abstract}


Adaptive model predictive control (MPC) robustly ensures safety while reducing uncertainty during operation. In this paper, a distributed version is proposed to deal with network systems featuring multiple agents and limited communication. 
To solve the problem in a distributed manner, structure is imposed on the control design ingredients \resp{without sacrificing performance.}
Decentralized and distributed adaptation schemes that allow for a reduction of the uncertainty online compatibly with the network topology are also proposed. The algorithm ensures robust constraint satisfaction, recursive feasibility and finite gain $\ell_2$ stability, and yields lower closed-loop cost compared to robust distributed MPC in simulations. 
\end{abstract}


\section{Introduction}

In the increasing number of applications involving complex interconnected networks, advanced control strategies are required to ensure safety while coping with coupled dynamics and constraints. A candidate is model predictive control (MPC), where a model of the system is used to optimize over state and control trajectories and guarantee constraint satisfaction \cite{rawlings2009model}. 
Robust MPC methods take into account uncertainty in the model, but they might lead to conservative actions when the uncertainty is large.
 Adaptive MPC (AMPC) is a technique which ensures robustness while updating the uncertainty using online identification (e.g. see \cite{lorenzen2019,parsi2020active,bujarbaruah2020adaptive}). 
In AMPC, uncertainty is captured by a parameteric state space model. Uncertainty bounds are updated using set membership identification \cite{milanese1991}, while a tube MPC approach ensures robust constraint satisfaction. Two current limitations motivated this work, and are addressed by the proposed algorithms.  The first is that the optimization problem which is solved online grows combinatorially with the state dimension. 
Although this was partially ameliorated in \cite{kohler2019linear}, the number of constraints still grows exponentially with the number of uncertain parameters
The second is that AMPC uses a centralized optimization problem and identification scheme, which cannot be applied to interconnected networks without a central unit.  

Centralized control of interconnected systems may suffer \resp{loss of privacy, lack of robustness against failure and high computational cost}. Thus, distributed control has become a strong contender for these systems \cite{antonelli2013interconnected}. In particular, distributed MPC (DMPC) has gained attention due to its ability to consider constraints \cite{christofides2013distributed}. In \resp{many} DMPC schemes, a network is decomposed into smaller subsystems. Each subsystem solves a local optimization problem iteratively while sharing information with other subsystems. This yields distributed optimal control problems that can be solved using distributed optimization (e.g. see \cite{boyd2011distributed}).
Various efforts have been devoted to developing robust DMPC schemes for uncertain systems \resp{(e.g. see \cite{shalmani,madistributed}). However, } 
none of the schemes consider online identification of the network’s parameters.

In this work, we propose a distributed AMPC (DAMPC) algorithm suitable for control of interconnected networks. The key observation is that, for interconnected networks, the state space matrices and constraints are sparse and structured, and this is leveraged to modify the AMPC optimization problem. 
The optimization parameters and variables are structured such that the problem can be solved using distributed optimization, \resp{while reducing the performance loss.} The number of variables and constraints in the resulting problem increases only linearly with the number of agents in the network. The MPC optimization problem is solved in a distributed fashion using alternating direction method of multipliers (ADMM) \cite{boyd2011distributed}.
\resp{Another novel contribution of this paper is the proposal of identification schemes for an interconnected network with no central unit, which has not been studied much in literature.}
Specifically, two different set membership algorithms, decentralized and distributed, are proposed. In the former, the agents exchange only the state measurements with their neighbors. In the latter, the bounds of shared parameters are also exchanged with the neighboring agents, leading to better performance. The resulting MPC algorithm ensures robust constraint satisfaction, recursive feasibility and closed loop stability of the system. \resp{ To the best of our knowledge, this is the first DMPC algorithm that guarantees robustness to parameteric uncertainty and performs online identification.}
We show the features of the proposed algorithms using an uncertain mass spring damper system.



\underline{Notation:} 
The Minkowski sum operator is denoted by $\oplus$. 
The matrix $O_{ij}$ represents a zero matrix with $i$ rows and $j$ columns, and $||x||_Q^2$ denotes $x^\top Q x$. The $ i ^\text{th}$ row of a matrix $ A $ is denoted by $ [A]_{i} $, and $\mathbb{N}_{1}^{a}$ denotes the set of integers from $1$ to $a$. 

\section{Adaptive MPC}\label{Background}
In this section, the robust adaptive model predictive control (AMPC) algorithm proposed in \cite{lorenzen2019} is briefly described.
Consider a discrete-time, linear time-invariant system with state $x_k \in \mathbb{R}^n$, control input $u_k \in \mathbb{R}^m$ and disturbance $w_k$ at the time step $k$, and system dynamics 
\begin{equation} \label{eq:Dynamics}
    x_{k+1} = A(\theta) x_k + B(\theta) u_k + w_k,
\end{equation}
where $\theta \in \mathbb{R}^p$ is a constant parameter and its true value $\theta^*$ is unknown. The state matrices are parameterized as
\begin{align} \label{eq:Parameterization}
\begin{split}
    A(\theta) = A_0 + \displaystyle\sum_{i=1}^{p} A_i [\theta]_i, \quad  B(\theta) = B_0 + \displaystyle\sum_{i=1}^{p} B_i [\theta]_i,
\end{split}
\end{align}
where  $\theta$ belongs to the known bounded polytope $\Theta_0 {:=} \{\theta | H_{\theta} \theta {\le} h_{\theta_0} \},$ such that $ \theta^* {\in} \Theta_0$ and $ H_{\theta}{\in}\mathbb{R}^{n_{\theta}\times p}$. 
The disturbance $w_k$ lies in a known bounded polytope $
    \mathbb{W} $. The states and inputs of the system must satisfy the constraints described by the compact polytope $\mathbb{Z} = \left\{(x_k,u_k) \bigr|  F x_k + G u_k \le \mathbf{1}\right\}$, where $F{\in}\mathbb{R}^{n_Z\times n}$ \resp{ and $G{\in}\mathbb{R}^{n_Z\times m}$}.
The objective is to regulate the system state from the initial condition $x_0$ to the origin, while satisfying the constraints \resp{for all possible realizations of uncertainty $\theta$ and disturbance $w_k$}.

Set-membership identification \cite{milanese1991} is performed to update the parameter set, which is given by $\Theta_k := \{\theta \in \mathbb{R}^p | H_{\theta} \theta \le h_{\theta_k} \}$. The bounds $h_{\theta_k}{\in}\mathbb{R}^{n_{\theta}}$ are updated online using a linear program. To ensure robustness against parameter uncertainties, a tube MPC approach is used \cite{langson2004}. The control input in the MPC optimization problem is parameterized as $ u_{l|k} =  K x_{l|k} + v_{l|k} $, where subscript $\{\cdot\}_{l|k}$ denotes the $l$ steps ahead prediction at time step $k$, $v_{l|k}$ is the MPC optimization variable  and  $K$ is a control gain stabilizing all the plants in $\Theta_0$. A state tube is defined using a sequence of sets $\{\mathbb{X}_{l|k}\}$ satisfying
\begin{align}
\mathbb{X}_{0|k} \ni \{x_{k}\},  \quad
\mathbb{X}_{l+1|k} &\supseteq A(\theta)\mathbb{X}_{l|k}  \oplus B(\theta)u_{l|k} \oplus \mathbb{W} , \label{eq:SetInclusions}
\end{align}
for all $\theta {\in} \Theta_k$ and$ \: l {\in} \mathbb{N}_0^{N-1}$, \resp{where $N$ is the prediction horizon}. The sets $ \mathbb{X}_{l|k} $ are parameterized according to $\mathbb{X}_{l|k} = \{z_{l|k}\} \oplus \alpha_{l|k} \mathbb{X}_{0},$ where $z_{l|k}{\in}\mathbb{R}^n,\alpha_{l|k}{\in}\mathbb{R}_{\ge0}$ are optimization variables and $\mathbb{X}_0$ is designed offline using the matrix $H_x{\in}\mathbb{R}^{n_x\times n}$ as
\begin{align}\label{tube_0}
 \mathbb{X}_0 &:= \{x| H_x x \le \mathbf{1}\} = \text{co}\{x^{1},x^{2},\ldots,x^{q}\},
\end{align}
where $x^i, i\in\mathbb{N}_{1}^{q}$ are $q$ prespecified vertices and $\text{co}\{.\}$ denotes their convex hull. The states in the set $\mathbb{X}_{N|k}$ are constrained to lie in a terminal set $
\mathbb{X}_T := \{ x|H_x x\le\bar{\alpha} \mathbf{1} \},$ that is robustly invariant under the terminal controller $u=Kx$. \resp{The constant $\bar{\alpha}$ is computed offline as the maximum value of $\alpha$ for which $\alpha\mathbb{X}_0\in \mathbb{Z}$.}

In the MPC problem, the state and input constraints, set inclusions for the tube, and the terminal constraints are formulated in terms of the optimization variables $\{z_{l|k},v_{l|k},\alpha_{l|k}\}$. For simplicity, we define the notation
\begin{align}\label{eq:x_jlk}
x_{l|k}^{j} &= z_{l|k} + \alpha_{l|k} x^{j}, \quad &&d_{l|k}^{j} = A_0 x_{l|k}^{j} + B_0 u_{l|k}^{j} - z_{l+1|k},  \nonumber \\ 
u_{l|k}^{j} &= Kx_{l|k}^{j} + v_{l|k}, \quad &&D_{l|k}^{j} = D(x_{l|k}^{j},u_{l|k}^{j}) , 
\end{align}
where $j \in \mathbb{N}_{1}^{\resp{q}} ,l \in \mathbb{N}_{0}^{N-1} $, and the matrix $D(x_k,u_k) = [A_1 x_k+ B_1 u_k \: \ldots \: A_p x_k+ B_p u_k]$. The vectors $\bar{f},\bar{w}$ are precomputed offline such that $[\bar{f}]_{i} = \displaystyle\max_{x\in \mathbb{X}_0} [F+GK]_i x, \: i {\in} \mathbb{N}_{1}^{n_Z} $ and $ [\bar{w}]_{j} = \displaystyle\max_{w\in \mathbb{W}} \: [H_x]_j w , \:  j {\in} \mathbb{N}_{1}^{n_x} $.
A quadratic cost function is used in MPC following a certainty-equivalence approach. For this purpose, an estimate of the parameter $(\hat{\theta}_k)$ is obtained using a least mean squares filter \cite{hassibi1995lms} and projected onto the set $\Theta_k$ at each time step. 
The final MPC optimization problem can be written as 
\resp{
\begin{subequations}\label{eq:centralizedOptimizationProblem}
\begin{align}%
\displaystyle\min_{\substack{z_{l|k},v_{l|k}, \\\alpha_{l|k},\Lambda_{l|k}^j}}  \displaystyle\sum_{l=0}^{N-1} \left( ||\hat{x}_{l|k}||_Q^2 + ||\hat{u}_{l|k}||_R^2 \right)&+  ||\hat{x}_{N|k}||_P^2,  \\
\text{s.t.} \qquad \qquad -H_x z_{0|k} -\alpha_{0|k}\mathbf{1} &\le -H_x x_k ,\label{eq:MPCa}	\\
 (F+GK)z_{l|k} + Gv_{l|k} + \alpha_{l|k}\bar{f} &\le \mathbf{1},\label{eq:MPCb}\\
	\Lambda_{l|k}^{j} h_{\theta_k} + H_x d_{l|k}^{j} -\alpha_{l+1|k} \mathbf{1} &\le -\bar{w} ,\label{eq:MPCc}\\
	H_x D_{l|k}^{j} &= \Lambda_{l|k}^{j} H_{\theta},\label{eq:MPCd}\\
 \hat{x}_{0|k} = x_k, \quad A(\hat{\theta}_k) \hat{x}_{l|k} + B(\hat{\theta}_k) \hat{u}_{l|k} &=\hat{x}_{l+1|k}, \label{eq:MPCe}\\
  \quad  z_{N|k} = 0,\quad \alpha_{N|k} &\le \bar{\alpha},\label{eq:MPCf}
\end{align} 
\end{subequations}
}
 where $ j{\in} \mathbb{N}_{1}^{q} $, $ l{\in} \mathbb{N}_{0}^{N-1} $ and $P,Q,R$ are positive definite matrices. \resp{The state tube is initialized in \eqref{eq:MPCa} and the state and input constraints are reformulated in terms of the optimization variables in \eqref{eq:MPCb}. The state tube dynamics are propagated inside the prediction horizon in \eqref{eq:MPCc} and \eqref{eq:MPCd} using the Lagrange multiplier matrices $\Lambda_{l|k}^j {\in} \mathbb{R}^{n_x\times n_\theta}$, $ j{\in} \mathbb{N}_{1}^{q}, l{\in} \mathbb{N}_{0}^{N-1}$, which have non-negative entries.  The certainty equivalence trajectory is defined in \eqref{eq:MPCe} and the terminal constraints are shown in \eqref{eq:MPCf}.} Note that the number of variables and constraints in \eqref{eq:centralizedOptimizationProblem} depends on $q$, i.e., the number of vertices in $\mathbb{X}_0$ (\ref{tube_0}), which increases combinatorially with the state dimension.

\section{Distributed adaptive MPC}\label{Formulation}
We present here a DAMPC algorithm which enables a network of agents to be controlled using an adaptive MPC approach. The algorithm requires no central unit for online implementation. Moreover, the size of the MPC optimization problem only linearly grows with the number of agents in the network.

Consider an interconnected system of $S$ agents (also referred to as subsystems). Each agent has uncertain parameters, some of which might be common to multiple agents. The goal is to impose structure in the design matrices and sets defined in Section \ref{Background} so that \eqref{eq:centralizedOptimizationProblem} can be solved in a distributed manner. 
As a preamble, note that variables and matrices corresponding to agent $s$ are denoted using the subscript, for example $A_{0,s}$, $ v_{l|k,s}$. Then, agent $s$ in the interconnected system obeys
\begin{equation}\label{eq:DistributedDynamics}
    x_{k+1,s} = A_{\text{d},s}(\theta_{s}) x_{k,\pazocal{N}_s} + B_{\text{d},s}(\theta_{s}) u_{k,s} + w_{k,s} ,
\end{equation}
where the subscript d is used to differentiate the matrices $A_{\text{d},s}$ from $A_{i}$ in \eqref{eq:Parameterization}. The \resp{set of neighbors $\pazocal{N}_s$} contains all the agents coupled dynamically or through the constraints with agent $s$, including itself. The vector $x_{k,\pazocal{N}_s}$ comprises states of all the agents in the set $\pazocal{N}_s$, and $\theta_{s}{\in} \mathbb{R}^{p_{s}}$ consists of parameters affecting agent $s$.  It is assumed that agents can communicate with all of their neighbors, and that \resp{their inputs are} decoupled.



\begin{figure}[t]
    \vspace{0.05cm}
    \centering
    \begin{tikzpicture}[yscale=0.4]
		
		\draw[thick](0,0) rectangle (1,1);
		\node at (0.5,0.5){$ S_1 $};
		\node[anchor=north] at (0.5,0){$ [\theta]_1 $};
		
		\coordinate (v1) at (1.5,0);
		\draw[thick] (1,0.5) -- ($ (v1) + (0,0.5) $);
		\draw[thick]($ (v1) + (0,0) $) rectangle ($ (v1) + (1,1) $);
		\node at ($ (v1) + (0.5,0.5) $){$ S_2 $};
		\node[anchor=north] at ($ (v1) + (0.5,0) $){$ [\theta]_1,[\theta]_2$};
		
		\coordinate (v2) at (3.0,0);
		\draw[thick] ($ (v1) + (1,0.5) $) -- ($ (v2) + (0,0.5) $);
		\draw[thick]($ (v2) + (0,0) $) rectangle ($ (v2) + (1,1) $);
		\node at ($ (v2) + (0.5,0.5) $){$ S_3 $};
		\node[anchor=north] at ($ (v2) + (0.5,0) $){$ [\theta]_2,[\theta]_3$};
	
		\coordinate (v3) at (4.5,0);
		\draw[thick] ($ (v2) + (1,0.5) $) -- ($ (v3) + (0,0.5) $);
		\draw[thick]($ (v3) + (0,0) $) rectangle ($ (v3) + (1,1) $);
		\node at ($ (v3) + (0.5,0.5) $){$ S_4 $};
		\node[anchor=north] at ($ (v3) + (0.5,0) $){$ [\theta]_3$};
		
	\end{tikzpicture}
    \setlength{\belowcaptionskip}{-10pt}
    \caption{Example interconnected network. The parameters affecting each agent are reported below.}
    \label{fig:NetworkExample}
\end{figure}
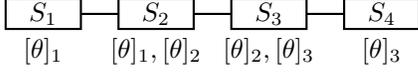

For the sake of clarity, a notional network topology is used to illustrate the structure imposed on the MPC problem. Consider the 4-agent network shown in Figure \ref{fig:NetworkExample}, with each agent having three states and two inputs. For this example, the state matrix has the structure
\begin{gather*}
A(\theta) = \begin{bmatrix}
  \tikzmark{Al1} & \tikzmark{Ar1}  & O_{33} & O_{33}\\
  \tikzmark{Al2} &   & \tikzmark{Ar2} & O_{33}\\
  O_{33} &  \tikzmark{Al3} &  & \tikzmark{Ar3} \\
 O_{33} &  O_{33} & \tikzmark{Al4} & \tikzmark{Ar4} 
\end{bmatrix},
\end{gather*}
\resp{while the input matrix $B(\theta)$ has a block structure connecting each agent's inputs to its own states.} We assume that the constraints to be satisfied by the agent $s$ are given by the compact polytope $\mathbb{Z}_{s} = \left\{(x_{k,\pazocal{N}_s},u_{k,s}) \bigr|  F_{s} x_{k,\pazocal{N}_s} + G_{s} u_{k,s} \le \mathbf{1}\right\}.$
Recall that states of non-neighboring agents are not coupled by constraints, and the input constraints are fully decoupled. 
\DrawBox{Al1}{Ar1}{\textcolor{black}{$A_{\text{d},1}(\theta_1)$}}
\DrawBox{Al2}{Ar2}{\textcolor{black}{$A_{\text{d},2}(\theta_2)$}}
\DrawBox{Al3}{Ar3}{\textcolor{black}{$A_{\text{d},3}(\theta_3)$}}
\DrawBox{Al4}{Ar4}{\textcolor{black}{$A_{\text{d},4}(\theta_4)$}}
\vspace{-0.1cm}

\subsection{Structured design matrices}\label{sec:Structure}
The control gain $K$ is designed such that the feedback can be implemented using a distributed feedback controller. For this purpose, in each row of $K$ corresponding to agent $s$, only the columns corresponding to the set $\pazocal{N}_s$ can be nonzero. These values can be extracted to obtain the control gain $K_{s}$ for each agent, and the MPC control input satisfies $u_{l|k,s} = K_{s}x_{l|k,\pazocal{N}_s} + v_{l|k,s}$. The structured gain $K$ can be computed offline, for example as suggested in \cite{kothare1996,conte2016distributed}. \resp{Although \cite{kothare1996} is tailored for centralized MPC, this method can be extended to distributed systems by imposing appropriate structure on the decision variables. Furthermore, although \cite{conte2016distributed} is proposed for nominal MPC, it can be extended to 
the uncertain system \eqref{eq:Dynamics} by ensuring that the corresponding constraints are satisfied for all vertices of $\Theta_0$.} For the example in Figure \ref{fig:NetworkExample}, the structure of $K$ is 
\begin{equation}\label{eq:Kstructure}
    K =  \begin{bmatrix}
  \tikzmark{Kl1} & \tikzmark{Kr1}  & O_{23} & O_{23}\\
  \tikzmark{Kl2} &   & \tikzmark{Kr2} & O_{23}\\
  O_{23} &  \tikzmark{Kl3} &  & \tikzmark{Kr3} \\
 O_{23} &  O_{23} & \tikzmark{Kl4} & \tikzmark{Kr4} 
\end{bmatrix}.
\end{equation}
\DrawBox{Kl1}{Kr1}{\textcolor{black}{$K_{1}$}}
\DrawBox{Kl2}{Kr2}{\textcolor{black}{$K_{2}$}}
\DrawBox{Kl3}{Kr3}{\textcolor{black}{$K_{3}$}}
\DrawBox{Kl4}{Kr4}{\textcolor{black}{$K_{4}$}}
\vspace{-0.1cm} 

The matrix $H_x$ defined in \eqref{tube_0} is structured so that, in each row, only the columns corresponding to a single agent can be nonzero. This results in a block matrix structure, where each block represents the structure of the state tube of the corresponding agent as
\begin{align}\label{eq:X_0s}
 \mathbb{X}_{0,s} &:= \{x_s| H_{x,s} x_s \le \mathbf{1}\} = \text{co}\{x_s^{1},x_s^{2},\ldots,x_s^{q_s}\}.
\end{align}
The structure \eqref{eq:X_0s} implies that the number of vertices to be evaluated depends on $\sum_{s=1}^{S}q_s$ instead of $q$, resulting in linear growth in the number of variables and constraints with the number of agents. For the example shown in Figure \ref{fig:NetworkExample}, the structure of $H_x$ is
\begin{equation}\label{eq:Hx_structure}
    H_x =  \begin{bmatrix}
  \tikzmark{Hxl1} {\small H_{x,1}}  \tikzmark{Hxr1} & O_{83}  & O_{83} & O_{83}\\
     O_{83} &\tikzmark{Hxl2} {\small H_{x,2}} \tikzmark{Hxr2} & O_{83} & O_{83}\\
  O_{83} & O_{83}&  \tikzmark{Hxl3} {\small H_{x,3}} \tikzmark{Hxr3} & O_{83}\\
 O_{83} &  O_{83}& O_{83} & \tikzmark{Hxl4} {\small H_{x,4}} \tikzmark{Hxr4} 
\end{bmatrix},
\end{equation}
where each agent is using 8 constraints to define $X_{0,s}$. The computation of a structured $H_x$ while ensuring  robust invariance of the terminal set is a difficult task, and is the topic of future research. The block structure \eqref{eq:X_0s} enables an additional degree of freedom in parameterizing the state tube according to $\mathbb{X}_{l|k,s} = z_{l|k,s} \oplus \alpha_{l|k,s}\mathbb{X}_{0,s}$, where $\alpha_{l|k,s}$ chosen by each subsystem need not be equal.

The matrix $H_{\theta}$ is structured so that different parameters can have a nonzero coefficient in the same hyperplane only if they affect a common agent. The structure chosen in $H_{\theta}$ affects the Lagrange multipliers $\Lambda^j_{l|k}$ used to build the state tube. If \resp{a constraint affects} multiple agents, they must use the same Lagrange multipliers, resulting in a large number of shared variables. To solve this problem, redundant constraints are added to $\{H_{\theta},h_{\theta_k} \} $, and denoted as $\{\Tilde{H}_{\theta},\Tilde{h}_{\theta_k} \}$.  These constraints are formed by repeating constraints affecting multiple agents by the number of agents affected. The resulting matrix $\Tilde{H}_{\theta}$ has a block representation, where each block $\Tilde{H}_{\theta,s}$ and the corresponding terms $\Tilde{h}_{\theta_k,s}$ represent all the constraints on the parameters affecting agent $s$.  For the example in Figure \ref{fig:NetworkExample}, let the matrix $H_{\theta}$ be such that the constraints define upper and lower bounds on the parameters $[\theta]_1,[\theta]_2,[\theta]_3$, and two constraints jointly on the parameters $[\theta]_1,[\theta]_2$. This results in the structure
\begin{equation*}
{H}_\theta  {=}  \begin{bmatrix}
  \tikzmark{htl1} {\small H_{\theta \mbox{-} 1}}  \tikzmark{htr1} & O_{21}  & O_{21} \vspace{0.05in} \\ 
  \tikzmark{htl12}  & \tikzmark{htr12} & O_{21}\\
   O_{21} &  \tikzmark{htl2}   {\small H_{\theta \mbox{-} 2}} \tikzmark{htr2} & O_{21}\\
  O_{21} &  O_{21} & \tikzmark{htl3}  { H_{\theta \mbox{-} 3}}  \tikzmark{htr3}
  \end{bmatrix}, \:
   \Tilde{H}_\theta  {= } \begin{bmatrix} 
  \tikzmark{Htl1} {\Tilde{H}_{\theta,1}}  \tikzmark{Htr1} & O_{21}  & O_{21} \\[6pt]
     \tikzmark{Htl2}  &  \tikzmark{Htr2} & O_{61}\\[3pt]
  O_{41} & \tikzmark{Htl3}  &   \tikzmark{Htr3} \\[3pt]
 O_{21} &  O_{21} & \tikzmark{Htl4} {\Tilde{H}_{\theta,4}} \tikzmark{Htr4} 
\end{bmatrix},
\end{equation*}
where the subscript $i,j$ in $H_{\theta \mbox{-} i,j}$ indicate the parameters involved in the constraints defined by the block.
\DrawBoxFive{Htl1}{Htr1}{}
\DrawBoxThree{Htl2}{Htr2}{${\Tilde{H}_{\theta,2}}$}
\DrawBoxThree{Htl3}{Htr3}{${\Tilde{H}_{\theta,3}}$}
\DrawBoxFive{Htl4}{Htr4}{}
\DrawBoxTwo{htl1}{htr1}{}
\DrawBoxSix{htl12}{htr12}{$ {H_{\theta \mbox{-} 1,2}}$}
\DrawBoxTwo{htl2}{htr2}{}
\DrawBoxTwo{htl3}{htr3}{}
\DrawBoxTwo{Hxl1}{Hxr1}{}
\DrawBoxTwo{Hxl2}{Hxr2}{}
\DrawBoxTwo{Hxl3}{Hxr3}{}
\DrawBoxTwo{Hxl4}{Hxr4}{}
Due to the additional rows in $\Tilde{H}_{\theta}$, additional columns are added to $\Lambda^j_{l|k}$ and the new matrix is denoted as $\Tilde{\Lambda}^j_{l|k}$. In $\Tilde{\Lambda}^j_{l|k}$, each row corresponding to $H_{x,s}$ has nonzero elements only in the columns corresponding to $H_{\theta,s}$. For the example system, this results in the structure 
\begin{equation}
    \Tilde{\Lambda}^j_{l|k} =  \begin{bmatrix}
  \tikzmark{Ll1} {\Lambda}^j_{l|k,1} \tikzmark{Lr1} & O_{86}  & O_{84} & O_{82}\\ 
     O_{82} &\tikzmark{Ll2} {\Lambda}^j_{l|k,2}\tikzmark{Lr2} & O_{84} & O_{82}\\
  O_{82} & O_{86}&  \tikzmark{Ll3} {\Lambda}^j_{l|k,3} \tikzmark{Lr3} & O_{82}\\
 O_{82} &  O_{86}& O_{84} & \tikzmark{Ll4} {\Lambda}^j_{l|k,4}\tikzmark{Lr4} 
\end{bmatrix}.
\end{equation}
\DrawBoxFour{Ll1}{Lr1}{}
\DrawBoxFour{Ll2}{Lr2}{}
\DrawBoxFour{Ll3}{Lr3}{}
\DrawBoxFour{Ll4}{Lr4}{}

Finally, the cost matrices $P,Q,$ and $R$ have a block diagonal structure, where the corresponding blocks for each agent are denoted as $P_s,Q_s,$ and $R_s$.

\subsection{Set membership identification}
To perform DAMPC, each agent defines the feasible parameter set $\Theta_{k,s}:=\{\theta_s|\Tilde{H}_{\theta,s}\theta_s\le \tilde{h}_{\theta_k,s} \} $ which is updated using set membership identification. \resp{However, the identification method proposed in \cite{lorenzen2019} cannot be used here, since it requires a central agent with access to all the measurements. We propose two approaches to update $\Theta_{k,s}$ with local knowledge. }
At each time step, the state measurements are shared with the neighbors, and used to construct a set of non-falsified parameters 
\begin{align} \label{eq:SimpleNonfalsified}
\Delta_{k,s} &:= \biggr\{
\theta_s \: \biggr|x_{k+1,s} {-} A_{\text{d},s}(\theta_s) x_{k,\pazocal{N}_s} {-} B_{\text{d},s}(\theta_s) u_{k,s} \in \mathbb{W}_{s},\biggr\} \nonumber\\ 
&= \left\{\theta_s \: |\: H_{\Delta,s} \theta_s \le h_{\Delta,s} \right\}.
\end{align}
The parameter set is then updated by computing $[\tilde{h}_{\theta_{k},s}]_i$ as \resp{the optimal value} of
\begin{align}\label{eq:Theta_k_LP}
\max_{\theta_s} \: [\Tilde{H}_{\theta,s}]_{i} \theta_s, \quad
 \text{s. t. } 
 \begin{bmatrix}
    \Tilde{H}_{\theta,s} \\ H_{\Delta,s}
\end{bmatrix} \theta_s
 \le \begin{bmatrix}
  \tilde{h}_{\theta_{k-1},s} \\ h_{\Delta,s}
 \end{bmatrix},
\end{align}
for $i\in \mathbb{N}_{1}^{n_\theta,s} $. Problem \eqref{eq:SimpleNonfalsified}-\eqref{eq:Theta_k_LP} provides a decentralized identification scheme, where agent $s$ updates its own parameters using the measurements of its neighbors. 
An alternative is distributed identification, where parameter bounds are also communicated with neighbors. This can be useful when multiple agents are affected by the same parameter uncertainty, but have different identification performance.
Algorithm \ref{Alg:Dist_ID} details the procedure for such a distributed identification scheme. As a prerequisite, the algorithm requires that the polytope $\Theta_{k,s}$ has hyperplanes orthogonal to each parameter's axis. The corresponding elements in $h_{\theta_k,s}$  are the upper ($[\theta^\text{ub}_{k,s}]_{i}$) and lower ($[\theta^\text{lb}_{k,s}]_{i}$) bounds of the $i^\text{th}$ parameter, as computed by agent $s$. In the first step of the algorithm, decentralized identification is performed to obtain $h_{\theta_k,s}$. Then, the bounds of the shared parameters are exchanged with the corresponding neighbors. The upper and lower bounds are then updated using all the information \resp{received} as 
\begin{align}\label{eq:ShareUpdateBounds}
    [\bar{\theta}_{k,s}]_{i} = \min_{j\in \pazocal{N}_s} ([\theta^\text{ub}_{k,j}]_{\hat{i}}), \quad 
    [\underline{\theta}_{k,s}]_i = \max_{j\in \pazocal{N}_s} ([\theta^\text{lb}_{k,j}]_{\hat{i}}),
\end{align}
where $[\theta^\text{ub}_{k,j}]_{\hat{i}},[\theta^\text{lb}_{k,j}]_{\hat{i}}$ denote the upper and lower bounds of the parameter $[{\theta}_{k,s}]_i$ as computed by the agent $j$. The values of $h_{\theta_k,s}$ are updated using $[\bar{\theta}_{k,s}]_{i},[\underline{\theta}_{k,s}]_i$. 
\setlength{\textfloatsep}{2pt}
\begin{algorithm}[tb]
    \vspace{0.05cm}
	\caption{Distributed identification}\label{Alg:Dist_ID} 
	\begin{algorithmic}[1]
        \State Perform decentralized identification \eqref{eq:SimpleNonfalsified}-\eqref{eq:Theta_k_LP}
		\State Communicate $\theta^\text{ub}_{k,s},\theta^\text{lb}_{k,s}$ with corresponding neighbors
		\State Compute the bounds according to \eqref{eq:ShareUpdateBounds}
		\State Update $h_{\theta_k,s}$ using $\bar{\theta}_{k,s}$ and $\underline{\theta}_{k,s}$.
	\end{algorithmic}
\end{algorithm}

\subsection{Distributed optimization}
Incorporating the structure described in Section \ref{sec:Structure}, the optimization problem \eqref{eq:centralizedOptimizationProblem} can be reformulated as

\resp{
\begin{align}\label{eq:distributedOptimizationProblem}
\displaystyle\min_{\substack{z_{l|k,s},v_{l|k,s}\\\alpha_{l|k,s},\tilde{\Lambda}_{l|k,s}^j}} \displaystyle\sum_{s=1}^{S}  \displaystyle\sum_{l=0}^{N-1} \bigr( ||\hat{x}_{l|k,s}||_{Q_s}^2 {+} ||\hat{u}_{l|k,s}||_{R_s}^2 &\bigr) {+} ||\hat{x}_{N|k,s}||_{P_s}^2  \nonumber \\
\text{s.t.} \qquad \qquad -H_{x,s} z_{0|k,s} -\alpha_{0|k,s}\mathbf{1} &\le -H_{x,s} x_{k,s} ,\nonumber \\
 F_{\text{cl},s}z_{l|k,s} + G_sv_{l|k,s} + \alpha_{l|k,s}\bar{f}_s &\le \mathbf{1},\nonumber\\
	\Tilde\Lambda_{l|k,s}^{j} \Tilde h_{\theta_k,s} {+} H_{x,s} d_{l|k,s}^{j} {-}\alpha_{l+1|k,s} \mathbf{1} &\le -\bar{w}_{s} ,\\
	H_{x,s} D_{l|k,s}^{j} &= \Tilde\Lambda_{l|k,s}^{j} \Tilde H_{\theta,s},\nonumber \\
  A_{s}(\hat{\theta}_{k,s}) \hat{x}_{l|k,\pazocal{N}_s} + B_{s}(\hat{\theta}_{k,s}) \hat{u}_{l|k,s} &=\hat{x}_{l+1|k,s},\nonumber \\
 \hat{x}_{0|k,s} = x_{k,s}, \quad  z_{N|k,s} = 0,\quad \alpha_{N|k,s} &\le \bar{\alpha}_{s}, \nonumber 
\end{align}
}
where $F_{\text{cl},s} {=} F_s{+}G_sK_s$,  $ j{\in} \mathbb{N}_{1}^{v_s} $, $ l{\in} \mathbb{N}_{0}^{N-1} , s{\in}\mathbb{N}_{1}^{S}$, and the constants $\bar{f}_s,\bar{w}_s,\bar{\alpha}_s$ are calculated offline. \resp{The parameter estimate $\hat{\theta}_{k,s}$ is computed by agent $s$ using a least mean squares filter. \resp{Note that the constraints in \eqref{eq:distributedOptimizationProblem} are the same as those in \eqref{eq:centralizedOptimizationProblem}, but with a distributed structure.}
}
\begin{pro}
The DAMPC algorithm guarantees robust constraint satisfaction and recursive feasibility of the optimization problem. In addition, the closed loop defined by the system \eqref{eq:Dynamics} and the DAMPC scheme is finite gain $\ell_2$ stable. 
\end{pro}
\resp{We provide only an outline of the proof that follows similar lines to those of Lemma 5 and Theorem 14 from \cite{lorenzen2019}}. \resp{In addition to the structure imposed on $K$ and $H_x$, which does not affect the proof, the proposed changes to the AMPC algorithm are: a distributed identification scheme; repeated constraints in the parameter set; and additional Lagrange multipliers.} The \resp{added} constraints and variables \resp{are redundant, and} do not affect the feasibility of the optimization problem. The proposed decentralized and distributed identification algorithms guarantee by design that $\theta^* {\in} \Theta_{k}$ and $\Theta_{k}{\subseteq}\Theta_{k-1}$, \resp{ and thus, recursive feasibility is preserved.

To ensure finite gain $\ell_2$ stability, there must exist constants $c_0,c_1,c_2 \ge0$ such that for $m\in\mathbb{N}_+$
\begin{align}\label{eq:MPCproof0}
    \sum_{s=0}^{S}\sum_{k=0}^{m} ||x_{k,s}||^2 \le &\sum_{s=0}^{S} \bigr( c_0||x_{0,s}||^2
    \\   &+c_1 ||\hat{\theta}_{0,s}-\theta^*||^2 + c_2\sum_{k=0}^{m} || w_{k,s}||^2 \bigr).  \nonumber
\end{align}
Denoting the optimal value of the cost function in \eqref{eq:distributedOptimizationProblem} as $V_k$, the change in its value can be bounded as
\begin{equation}\label{eq:MPCproof1}
    V_{k+1} - V_k \le \sum_{s=0}^{S} \bigr(-c||x_{k,s}||^2 + c_A||x_{k+1,s} - \hat{x}_{1|k,s}||^2 \bigr),
\end{equation}
using  $c,c_A\ge0$. The upper bound in \eqref{eq:MPCproof1} is computed by constructing a candidate trajectory at time step $k+1$, using the parameter estimates $\hat{\theta}_{k+1,s}$ and shifting the optimal solution $v_{l|k,s}^*$ by one time step. The last control input variable $v_{N|k+1,s}$ is set to 0. The second term in the upper bound is the one step ahead prediction error. The least mean squares filters 
guarantee that the sum of squared prediction error can always be bounded using the initial error in the parameter estimate and the sum of the squared disturbances. This has been proven in \cite{lorenzen2019} for a centralized filter, and the bound is also valid for the DAMPC algorithm although it uses multiple copies of some parameters and a distributed filter. 
It can also be shown that $V_k$ can be upper bounded in the feasible set by a quadratic function of the states $x_{k,s}$. This means that summing \eqref{eq:MPCproof1} over $k$  results in \eqref{eq:MPCproof0}.
}

\begin{table}[]
    \centering    
    \vspace{0.05cm}
    \caption{ADMM variables computed by the second agent in the network example in Fig. \ref{fig:NetworkExample}} 
    \label{tab:ADMM}
    \begin{tabular}{|c|c|c|}
        \hline
        Step 1 & Step 2 & Step 3 \\
        \hline
        $C_{\pazocal{N}_2}=\left[C_{\pazocal{N}_2,1}^\top \ C_{\pazocal{N}_2,2}^\top \ C_{\pazocal{N}_3,2}\right]^\top$ & $T_2$ & $Y_{\pazocal{N}_2}$ \\
        $C_{\pazocal{N}_2,s}=\left[z_{l|k,s}^{(2)\top},\alpha_{l|k,s}^{(2)\top},\hat{x}_{l|k,s}^{(2)\top}\right]^\top_{\left\{\begin{matrix}  s \in \mathbb{N}_1^3, \\ l \in \mathbb{N}_0^{N-1}\end{matrix}\right\}}$ &  &  \\
        $E_2=\left[v_{l|k,2}^{(2)\top},\operatorname{vec}(\tilde{\Lambda})_{l|k,2}^{(2)\top},\hat{u}_{l|k,2}^{(2)\top}\right]^\top_{\{l \in \mathbb{N}_0^{N-1}\}}$ & & \\
        \hline
    \end{tabular}
\end{table}

We now cast the distributed optimal control problem \eqref{eq:distributedOptimizationProblem} into the general-form consensus problem 
\resp{in \cite[Section~7.2]{boyd2011distributed}}
so that it can be solved in a distributed fashion using ADMM. 
Note that two neighbours $s_1$ and $s_2$ share the decision variables $z_{l|k,s_1}$, $z_{l|k,s_2}$, $\alpha_{l|k,s_1}$, $\alpha_{l|k,s_2}$, $\hat{x}_{l|k,s_1}$ and $\hat{x}_{l|k,s_2}$. Hence, we define $C_{\pazocal{N}_s}$ to be a concatenated vector comprising local copies of all decision variables that subsystem $s$ shares with its neighbours (i.e. $C_{\pazocal{N}_s}=\left[z_{l|k,\sigma}^{(s)\top},\alpha_{l|k,\sigma}^{(s)\top},\hat{x}_{l|k,\sigma}^{(s)\top}\right]^\top_{\{l \in \mathbb{N}_0^{N-1}, \sigma \in \pazocal{N}_s\}}$ where the superscript $(\cdot)^{(s)}$ denotes the local copy computed by subsystem $s$). We also define $Y_{\pazocal{N}_s}$ as the vector of dual variables whose size is the same as that of $C_{\pazocal{N}_s}$. For any $\sigma \in \pazocal{N}_s$, we define the vector $C_{\pazocal{N}_s,\sigma}=\left[z_{l|k,\sigma}^{(s)\top},\alpha_{l|k,\sigma}^{(s)\top},\hat{x}_{l|k,\sigma}^{(s)\top}\right]^\top_{\{l \in \mathbb{N}_0^{N-1}\}}$ which is a subvector in $C_{\pazocal{N}_s}$.
On the other hand, a subsystem $s$ does not share the decision variables $v_{l|k,s}$, $\tilde{\Lambda}_{l|k,s}$ and $\hat{u}_{l|k,s}$. Thus, we define $E_s$ to be a concatenated vector comprising all decision variables that are not shared by subsystem $s$ (i.e. $E_s=\left[v_{l|k,s}^{(s)\top},\operatorname{vec}(\tilde{\Lambda})_{l|k,s}^{(s)\top},\hat{u}_{l|k,s}^{(s)\top}\right]^\top_{\{l \in \mathbb{N}_0^{N-1}\}}$ where  $\operatorname{vec(\cdot)}$ is the matrix vectorization operator).
Finally, we define $T$ as a concatenated vector comprising a global copy of all shared decision variables in the network. We denote the subvectors of $T$ corresponding to $C_{\pazocal{N}_s}$ and $C_{\pazocal{N}_s,s}$ by $T_{\pazocal{N}_s}$ and $T_s$, respectively. Note that for any $s \in \mathbb{N}_1^S$ and $\sigma \in \pazocal{N}_s$, the vectors $C_{\pazocal{N}_s,s}$ and $C_{\pazocal{N}_\sigma,s}$ correspond to the same subvector in $T$, that is, $T_s$.

\resp{By defining the local function $f_s(E_s,C_{\pazocal{N}_s})$ which encodes the local objective function and constraints of subsystem $s$ through indicator functions, the general-form consensus problem is given by}
\begin{equation}
    \begin{aligned}
        & \min_{E_s,C_{\pazocal{N}_s}} \sum_{s=1}^{S} f_s(E_s,C_{\pazocal{N}_s}) \ 
        s.t. \ C_{\pazocal{N}_s}=T_{\pazocal{N}_s} \  \forall s \in \mathbb{N}_1^S. \\
    \end{aligned}
\end{equation}
According to \cite{boyd2011distributed}, the ADMM steps performed by subsystem $s$ in iteration $t$ are given by
\begin{equation*}
    \{E_s^{t+1},C_{\pazocal{N}_s}^{t+1}\} = \argminA_{E_s,C_{\pazocal{N}_s}} 
    \left( 
    \begin{aligned}
        & f_s(E_s,C_{\pazocal{N}_s}) +Y_{\pazocal{N}_s}^{t^\top}C_{\pazocal{N}_s} \\
        & \quad \quad \quad +\frac{\rho}{2}\norm{C_{\pazocal{N}_s}-T_{\pazocal{N}_s}}_2 \\
    \end{aligned} 
    \right),
\end{equation*}
\begin{equation}
    \label{ADMM_steps}
    T_s^{t+1} = \frac{1}{|\pazocal{N}_s|}\sum_{\sigma \in \pazocal{N}_s} C_{\pazocal{N}_\sigma,s}^{t+1}, 
\end{equation}
\begin{equation*} 
        Y_{\pazocal{N}_s}^{t+1} = Y_{\pazocal{N}_s}^t + \rho (C_{\pazocal{N}_s}^{t+1}-{T}_{\pazocal{N}_s}^{t+1}),
\end{equation*}
where $\rho$ is a tuning scalar and $|\cdot|$ is the set cardinality operator. 
\resp{The ADMM algorithm \eqref{ADMM_steps} updates iteratively, for all $s \in \pazocal{N}_1^S$, the primal variables $C_{\pazocal{N}_s}$ and $E_s$, the global copies of all shared decision variables $T_s$, and the dual variables $Y_{\pazocal{N}_s}$, in a sequential manner.}
Although the vector $T$ is a global copy of the shared variables, the update of its subvectors $T_s$ in the second ADMM step depends only on local information given by subsystem $s$ and its neighbours. \resp{We also note that a few parameters, e.g. $\rho$ and $N$, should be known by all subsystems, but these are constant and hence can be communicated in the offline phase as in \cite{aboudonia}.} 
For the sake of clarity, Table \ref{tab:ADMM} shows the decision variables computed by the second agent in each ADMM step for the network in Fig. \ref{fig:NetworkExample}. 

\resp{
\begin{figure}[t]
    \vspace{0.05cm}
    \centering
    \begin{tikzpicture}[yscale=0.4]
		
		\draw[thick](0,0) rectangle (1,1);
		\node at (0.5,0.5){$ S_1 $};
		\node[anchor=north] at (0.5,0){$ k_{12} $};
		
		\coordinate (v1) at (1.5,0);
		\draw[thick] (1,0.5) -- ($ (v1) + (0,0.5) $);
		\draw[thick]($ (v1) + (0,0) $) rectangle ($ (v1) + (1,1) $);
		\node at ($ (v1) + (0.5,0.5) $){$ S_2 $};
		\node[anchor=north] at ($ (v1) + (0.5,0) $){$ k_{12},k_{23}$};
		
		\coordinate (v2) at (3.0,0);
		\draw[thick] ($ (v1) + (1,0.5) $) -- ($ (v2) + (0,0.5) $);
		\draw[thick]($ (v2) + (0,0) $) rectangle ($ (v2) + (1,1) $);
		\node at ($ (v2) + (0.5,0.5) $){$ S_3 $};
		\node[anchor=north] at ($ (v2) + (0.5,0) $){$k_{23},k_{34}$};
	
		\coordinate (v3) at (4.5,0);
		\draw[thick] ($ (v2) + (1,0.5) $) -- ($ (v3) + (0,0.5) $);
		\draw[thick]($ (v3) + (0,0) $) rectangle ($ (v3) + (1,1) $);
		\node at ($ (v3) + (0.5,0.5) $){$ S_4 $};
		\node[anchor=north] at ($ (v3) + (0.5,0) $){$k_{34},k_{45}$};
		
		\coordinate (v4) at (6.0,0);
		\draw[thick] ($ (v3) + (1,0.5) $) -- ($ (v4) + (0,0.5) $);
		\draw[thick]($ (v4) + (0,0) $) rectangle ($ (v4) + (1,1) $);
		\node at ($ (v4) + (0.5,0.5) $){$ S_5 $};
		\node[anchor=north] at ($ (v4) + (0.5,0) $){$k_{45}$};
	\end{tikzpicture}
    \setlength{\belowcaptionskip}{-1pt}
    \caption{\resp{Mass-spring-damper interconnected system.}}    
    \label{fig:SpringMassExample}
\end{figure}
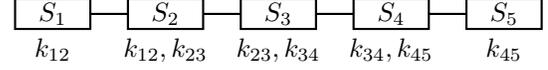
}

\section{Illustrative example}\label{Results}
We consider an interconnected mass-spring-damper system, consisting of five masses,  connected with four springs and four dampers, \resp{as shown in Figure \ref{fig:SpringMassExample}. }
All masses are equal to 1, and the damping coefficients are $\{2,2,2,1.5\}$. The true spring constants are unknown, but lie within a bounded set. The upper and lower bounds of each spring constant are given as $\{2.4,3.6,2.5,2.7\} {\pm} \kappa\{0.7,1,1,0.7\} $, where $\kappa$ is a model uncertainty scaling parameter. Each mass can be manipulated by a force, which is the control input. The dynamics are discretized using Euler discretization with a sampling time of 0.1s, with the position and velocity of each mass as the states of the system. The code used for the analyses of this example can be found \resp{in the repository\footnote{https://gitlab.ethz.ch/aparsi/distributedampcpublic}}. The system starts from the initial state $[2,1,2,-1,0,0,2,-1,2,1]^\top$. All states and inputs must lie within the bounds $[-5,5]$, \resp{and $w_k$ is bounded by $\pm 0.05$}. The MPC cost is defined by matrices $Q{=}I_{10},R{=}5I_5,P{=}100I_{10}$, where $I_n$ is an $n{\times}n$ identity matrix. 
The DAMPC scheme is designed with an horizon $N{=}5$. The gain matrix $K$ is obtained by imposing structure in the procedure described in \cite{kothare1996}. \resp{This results in a decentralized structure, where the control input of an agent is a function of only its states}. The state tube matrix $H_x$ is constructed by circumscribing the stabilized ellipsoidal region in the subspace defined by each agent's states using a polytope. The matrix $H_{\theta}$ is defined by upper and lower bounds on each parameter. 
The ADMM algorithm is implemented with $\rho{=}25$ for 400 iterations.

\begin{table}[t]
    \caption{Average reduction in closed loop cost achieved by adaptation (with both distributed and decentralized identification) for different sizes of model uncertainty.}
    \label{tab:Improvement}
    \centering
    \begin{tabular}{|c|c|c|c|c|c|c|c|}
    \hline
    $\kappa$ & 0.3 & 0.5 & 0.6 & 0.7 & 0.8 & 0.9 & 1.0 \\
    \hline
    DRMPC Cost & 376 & 417 & 462 & 527 & 610 & 730 & 1003 \\
    \hline
    $\%$ decrease & & & & & & & \\
    decentralized & 0.0 & 1.2 & 2.7 & 4.0 & 5.6 & 6.8 & 9.5 \\
    \hline
    $\%$ decrease & & & & & & & \\
    distributed & 0.1 & 2.1 & 4.4 & 6.6 & 9.3 & 11.3 & 15.1 \\
    \hline
    \end{tabular}
\end{table}

\begin{figure}[t]
    \centering
    \includegraphics[height=5cm]{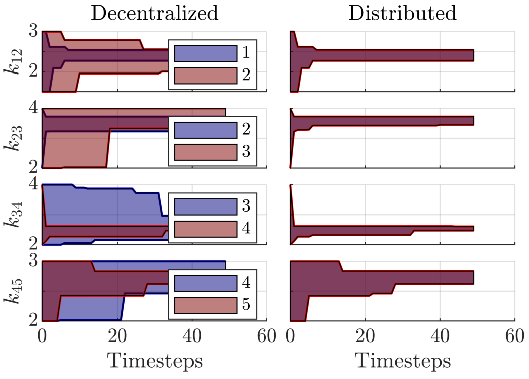}
    \caption{ 
    Parameter bounds obtained from decentralized and distributed identification techniques. On the left, estimates of each agent are indicated in the legend. On the right, shared estimates are shown in purple.} 
    \label{fig:CompareID}
\end{figure}

The closed loop performance of the DAMPC with decentralized and distributed identification schemes is compared to a distributed robust MPC (DRMPC) controller in Table \ref{tab:Improvement}. The simulations are performed for increasing values of $\kappa$, and the costs are averaged over 25 realizations of the disturbance $w_k$. The DRMPC scheme is obtained by using the same design matrices as in DAMPC but not performing adaptation. The reduction in the cost achieved by using the two proposed identification schemes is also shown in the last two \resp{rows}. For low values of $\kappa$, the performance of  the controllers is almost identical. As $\kappa$ increases, the closed loop cost is reduced when DAMPC schemes are used. 
The relative improvement in performance is higher when the model uncertainity is large. Moreover, distributed identification yields markedly larger cost reductions.
To provide more insights on this, Figure \ref{fig:CompareID} compares the performance of the proposed identification schemes. In decentralized identification, parameter bounds are updated separately by each agent. This leads to conservatism, especially in agents affected by multiple uncertainties. For example, agent 1 identifies $k_{12}$ better than agent 2, as it is only affected by one uncertainty. The updated bounds are not shared with agent 2, resulting in larger uncertainty for $k_{12,2}$. This problem is mitigated using distributed identification, where parameter bounds are also shared. Note that the performance of distributed identification is better than the intersection of bounds obtained from decentralized identification \resp{(e.g. see $k_{23}$ in Figure \ref{fig:CompareID})}. 

\section{Conclusions}
A distributed version of the AMPC algorithm is presented for interconnected systems with parametric uncertainty. The proposed algorithm is obtained by imposing structure on design matrices, such that the optimization problem can be solved in a distributed manner. The size of the resulting distributed optimization problem grows linearly with the number of agents in the network. 
Distributed and decentralized identification algorithms are proposed 
such that no centralized computations are required.
The former, by leveraging communication between agents, results in tighter parameter bounds and lower closed loop costs. 
\resp{It is anticipated that applications such as 
power networks \cite{riverso} and DC microgrids \cite{tucci}, 
where the coupling may depend on uncertain shared parameters, could benefit significantly of this novel scheme. The practical demonstration of these advantages is left for future work.
} 

\setlength{\dbltextfloatsep}{6pt}

\bibliographystyle{IEEEtran}

\bibliography{biblio_DAMPC}

\end{document}